\documentclass[amsmath,amssymb,showpacs,draft,preprint,floatfix]{revtex4}
\usepackage{graphicx}
\usepackage{latexsym}
\usepackage{amsmath}
\usepackage{amsfonts}
\usepackage{amssymb}
\begin{document}

\title{Relation between the autocorrelation and Wigner functions}
\author{ H\'ector Manuel Moya-Cessa$^{1,2}$, Demetrios N. Christodoulides$^2$  }
\affiliation{$^1$Instituto Nacional de Astrof\'{\i}sica, \'Optica
y Electr\'onica, Calle Luis Enrique Erro No. 1, 72840 Santa
Mar\'{\i}a Tonantzintla, Pue. \\
$^2$CREOL/College of Optics, University of Central Florida,
Orlando, FL, USA }

\begin{abstract}
We show a simple mechanism to measure the Wigner function of a
harmonic oscillator. For this system we also show that
autocorrelation  and  Wigner functions are equivalent.
\pacs{\bf Applied Mathematics  Information Sciences 7 (3), 839-841
(2013).}
\end{abstract}
\maketitle Non classical states of ions \cite{Moya3a} and
cavity fields \cite{Rosenhouse} have been  produced recently in
experiments around the world
\cite{varios1,varios2,varios3,varios4,varios5}. Once a given
nonclassical sate has been produced, it is important to count with
mechanisms that allow us to measure them, making the gathering  of
such information a key problem in quantum mechanics. Among this
mechanisms one can count with the fact that the passage of atoms
through a cavity may indicate the photon statistics of the cavity
field \cite{Rosenhouse}. For instance, information about the position
or momentum allows us to look for non classicality of the system.
However, it is possible to obtain full information from a system
by measuring, not some of its observables, but directly the
density matrix \cite{Wineland,Haroche}, i.e. obtaining information
about all possible observables. One of the possible ways of
obtaining such information is via a quasiprobability distribution
function, that may be related to the density matrix by using the
equation \cite{Royer,Knight}
\begin{equation}
F(\alpha,s)=
\frac{2}{\pi(1-s)}\sum_{k=0}^{\infty}\left(\frac{s+1}{s-1}\right)^k
\langle \alpha,k|\rho|\alpha,k\rangle
\end{equation}%
with $s$ the quasiprobability function's parameter that indicates
which is the relevant distribution ($s=-1$ Husimi \cite{Husimi},
$s=0$ Wigner \cite{Wigner} and $s=1$ Glauber-Sudarshan
\cite{Glauber,Sudarshan} distribution functions), $\rho$ the
density matrix and the states $|\alpha,k\rangle$ are the so-called
displaced number states \cite{Oliveira}.

It is well known that the Glauber-Sudarshan $P$-function is highly
singular (note the term $s-1$ in the denominator). It may be used
to measure non-classicality of states \cite{Kim}. It is not the
purpose of the present contribution to discuss about when is this
function well-behaved, however, it may be used to write the
density in a (diagonal) coherent state basis

\begin{equation}
\rho= \frac{1}{\pi}\int d^2\alpha
P(\alpha)|\alpha\rangle\langle\alpha|=\frac{1}{\pi}\int d^2\alpha
F(\alpha,1)|\alpha\rangle\langle\alpha|,
\end{equation}

that may be used to derive Fokker-Planck equations (partial
differential equations) from master equations (equations that
involve superoperators) \cite{Risken}.

The (Husimi) $Q$-function may be obtained from (1) by taking
$s=-1$. In such a case, the only term that survives in the sum is
$k=0$, that allow us to write

\begin{equation}
Q(\alpha)=F(\alpha,-1)=
\frac{1}{\pi}\langle\alpha|\rho|\alpha\rangle.
\end{equation}

Moreover, besides applications in classical optics, it has been
shown that these phase space distributions can be expressed, in
thermofield dynamics, as overlaps between the state of the system
and {\it thermal} coherent states \cite{Agarwal}, that is probably
the reason by which, systems subject to decay may still be
"measured" \cite{Amaro}.

Wineland's \cite{Wineland} and Haroche's \cite{Haroche} groups
used the above expression to measure the Wigner function ($s=0$
case) of the quantized motion of an ion and the quantized cavity
field, respectively. It is somehow simple to obtain a
quasiprobability distribution function from experimental data from
the above equation as there is already there a direct recipe. Let
us write equation (1) as
\begin{equation}
F(\alpha,s)=
\frac{2}{\pi(1-s)}\sum_{k=0}^{\infty}\left(\frac{s+1}{s-1}\right)^k
\langle k|D^{\dagger}(\alpha)\rho D(\alpha)|k\rangle
\end{equation}%
where $D(\alpha)=\exp(\alpha a^{\dagger}-\alpha^*a)$, with $a$ and
$a^{\dagger}$ the annihilation and creation operators
respectively, is the Glauber displacement operator. Note that, in
order to obtain a quasiprobability distribution function we need
to do the following: displace the system by an amplitude $\alpha$
and then measure the diagonal elements of the displaced density
matrix.

Equation (4) may be rewritten as
\begin{equation}
F(\alpha,s)=
\frac{2}{\pi(1-s)}Tr\left\{\left(\frac{s+1}{s-1}\right)^{a^{\dagger}a}
D^{\dagger}(\alpha)\rho D(\alpha)\right\}.
\end{equation}%
 By using the commutation properties under the symbol
of trace, and the system in a {\it pure} state $|\psi\rangle$, the
above equation may be casted into
\begin{eqnarray}
F(\alpha,s)= \frac{2}{\pi(1-s)}Tr\left\{
D(\alpha)\left(\frac{s+1}{s-1}\right)^{a^{\dagger}a}
D^{\dagger}(\alpha)\rho\right\}=\frac{2}{\pi(1-s)}\langle \psi |
D(\alpha)\left(\frac{s+1}{s-1}\right)^{a^{\dagger}a}
D^{\dagger}(\alpha)|\psi\rangle
\end{eqnarray}

Consider now a displaced harmonic oscillator with frequency
$\omega$
\begin{equation}
H=\omega a^{\dagger}a +\beta a^{\dagger} +\beta^*a
\end{equation}
with $\beta$ the amplitude of the displacement. One can directly
write the evolved wave function as (we set $\hbar=1$)
\begin{eqnarray}
|\psi(t)\rangle=e^{-iHt}|\psi(0)\rangle =
D^{\dagger}(\beta/\omega)e^{-i\omega t
a^{\dagger}a}D(\beta/\omega)|\psi(0)\rangle
\end{eqnarray}
From equation (6) we may obtain the {\it autocorrelation function}
\cite{Eleuch}

\begin{eqnarray}
A(t)=\langle \psi(0)|\psi(t)\rangle =\langle \psi(0)|
D^{\dagger}(\beta/\omega)e^{-i\omega t
a^{\dagger}a}D(\beta/\omega)|\psi(0)\rangle
\end{eqnarray}
that is very similar to equation (6). In fact,if we choose
$t=\pi/\omega$ in the above equation, it produces a term
\begin{eqnarray}
e^{-i\pi a^{\dagger}a}=(-1)^{a^{\dagger}a},
\end{eqnarray}
that is essential in the production of the Wigner function (the
alternating term), so tha by setting $s=0$ in equation (6), the
Wigner and autocorrelation functions become proportional:
\begin{equation}
F(\beta/\omega,0)=W(\beta/\omega)=\frac{2}{\pi}A(\pi/\omega)
\end{equation}
which is not surprising as the Wigner function is the generating
function for all spatial autocorrelation functions of the wave
function \cite{Editors}.

Thus, an eigenstate of the harmonic oscillator, namely, a number
state $|n\rangle$, may be easily measured, simply by choosing as
initial state $|\psi(0)\rangle=|n\rangle$, and projecting it with
the same number state. This can be done for instance in cavity
QED, by writing the evolved wavefunction as a density matrix and
then measuring its diagonal elements by passing atoms through the
cavity \cite{Amaro}. Note however, that, for every
displacement of the harmonic oscillator, a single value of the
Wigner function is obtained. Therefore for the reconstruction of
the Wigner function it is necessary a big number of experiments in
order to fill the phase space up.

Note that such systems may be emulated in classical light
propagation through waveguide arrays \cite{Leija1,Leija2} due to
the analogy between linear lattices and the atom-field interaction
\cite{Crisp}. Therefore, experiments leading to measurements of
quasiprobability distribution functions may be easier to implement
in classical optical systems.

In conclusion, we have shown a simple method to reconstruct the
Wigner function for the harmonic oscillator and have shown that
for this system, the autocorrelation function is proportional to
the Wigner function.

\end{document}